\documentclass[]{interact}

\usepackage{epstopdf}% To incorporate .eps illustrations using PDFLaTeX, etc.
\usepackage[caption=false]{subfig}% Support for small, `sub' figures and tables

\usepackage[numbers,sort&compress]{natbib}% Citation support using natbib.sty
\usepackage{amsmath,amssymb,amsfonts}
\usepackage{mathrsfs}
\usepackage{chemformula}
\usepackage{calrsfs}
\usepackage{color}
\usepackage[normalem]{ulem}

\usepackage[parfill]{parskip} % looks nicer - AW 

\bibpunct[, ]{[}{]}{,}{n}{,}{,}% Citation support using natbib.sty
\renewcommand\bibfont{\fontsize{10}{12}\selectfont}% Bibliography support using natbib.sty

\theoremstyle{plain}% Theorem-like structures provided by amsthm.sty

\theoremstyle{definition}

\theoremstyle{remark}

\newcommand{\II}{\textbf{I}}
\newcommand{\red}[1]{\textcolor{black}{#1}}
\newcommand{\ared}[1]{\textcolor{black}{#1}}

\newcommand{\beq}{\begin{equation}}
\newcommand{\eeq}{\end{equation}}
\newcommand{\pfr}[2]{\ensuremath{\frac{\partial #1}{\partial #2}}}
\newcommand{\pfi}[2]{\ensuremath{{\partial #1}/{\partial #2}}}
\newcommand{\ep}{\epsilon}
\newcommand\Pec{\mbox{\textit{Pe}}}
\newcommand\Pra{\mbox{\textit{Pr}}}
\newcommand\Lew{\mbox{\textit{Le}}}
\newcommand{\vect}[1]{\mathbf{#1}}

\newcommand{\Xt}{\tilde X}

\begin{document}

\articletype{ARTICLE TEMPLATE}% Specify the article type or omit as appropriate

\title{Effects of thermal expansion on Taylor dispersion-controlled diffusion flames}

\author{
\name{Prabakaran Rajamanickam\textsuperscript{a} and Adam D. Weiss\textsuperscript{b}}
\affil{\textsuperscript{a}Department of Aerospace Engineering, Auburn University, Auburn, AL 36849, USA; \\ \textsuperscript{b}ATA Engineering, San Diego, La Jolla, CA 92128, USA}
}

\maketitle

\begin{abstract}
A theoretical analysis is developed to investigate the effects of gas expansion due to heat release on unsteady diffusion flames evolving in a pipe flow in which \ared{the} mixing of reactants is controlled by Taylor's dispersion processes \ared{thereby extending a previously} developed theory based on the thermo-diffusive model. It is first shown that at times larger than radial diffusion times, the pressure gradient induced by the gas expansion is, in the first approximation, small in comparison with the prevailing pressure gradient driving the flow, indicating that corrections to \ared{the} background velocity profile \red{are} small. The corrections to the velocity components along with the leading-order mixing variables such as \ared{the} concentrations, temperature and density are solved for a Burke-Schumann flame. Due to the dependence of the effective Taylor diffusion coefficients on the gas density, quantitative and sometimes qualitative departures in predictions from the thermo-diffusive model are observed. 
\end{abstract}

\begin{keywords}
Taylor dispersion; Burke-Schumann flame; non-unity Lewis number; Poiseuille flow; Thermal expansion
\end{keywords}

\section{Introduction}

In an earlier paper by Li\~{n}\'{a}n \textit{et. al}~\cite{linan2020taylor} (referred to hereafter as \II), the Burke-Schumann diffusion flame embedded in an unsteady mixing layer developing when a fuel tank discharges fuel along a pipe of radius $a$ initially filled with air, was investigated. For a dimensional time $t^*<0$ the fuel in the tank and the air in the pipe \ared{are} separated by a partition or valve at the pipe entrance, \ared{there being} no fluid motion. At $t^*=0$, the partition is removed and the gas mixture is ignited. The pressure difference between the entrance and the open end will cause fluid motion along the pipe. 

In the absence of any thermal expansion due to heat release as it was assumed in \II, the fluid motion in the fully developed region is \red{dictated} by the Poiseuille law. For times larger than radial diffusion times i.e., $t^*\gg a^2/D_{O\infty}$, where $D_{O\infty}$ is the diffusivity of oxygen at standard temperature (constant properties in the fuel/oxidizer stream is denoted with the subscript $\infty$), \red{the mixing of reactants are governed by Taylor's dispersion processes. An observer moving with the mean pipe velocity, $U$, sees that the mixing variables such as concentrations, temperature etc., show no dependences on the radial coordinate, but evolve axially in time, very much like the normal diffusional spreading. The molecular diffusion coefficients are now, however, replaced by the so-called Taylor diffusion coefficients, which are functions only of the molecular diffusion coefficients and the Peclet number $\Pec=Ua/D_{O\infty}$. These observations are correct as long as the fluid density remains constant. As such, within the thermo-diffusive approximation, it is shown in \textbf{I} that the important flame characteristics (e.g. flame temperature and burning rate) depended on the fuel Lewis number and Peclet number only in a specific combination there defined as the ``effective" Lewis number. }

\ared{Here, we will extend the investigation carried out in \textbf{I} to include the effects of thermal expansion. We will show that the reduced parametric dependence of the solution on the ``effective" Lewis number, characteristic of the thermodiffusive approximation, no longer holds. It should be noted here that effects of thermal expansion have been studied previously~\cite{pearce2014taylor,daou2018taylor} in the context of steadily propagating premixed flames in narrow pipes and channels. There, asymptotic solutions for premixed flames were derived in the limit that the flame thickness becomes large compared with the radius of the pipe. In such cases} \red{the Taylor diffusion coefficients depend on the gas density also. As a result, the structure and the propagation speed of the flame depend significantly on the flow properties.} \ared{An analogous approach is taken here -- and previously in \textbf{I} -- with the thick-flame limit \cite{daou2002thick} for the premixed flame replaced by a thick mixing-layer limit for the diffusion flame in which the mixing of non-premixed reactants and products occurs over distances large compared to the pipe radius. As shown in \textbf{I}, the resulting analysis must account for a thin region encompassing the reaction sheet separating regions of fuel and oxidizer especially when preferential diffusion is present.}

%In this paper, we extend the investigation carried out in \textbf{I} where an asymptotic solution to the problem was developed in the limit $t^*\gg a^2/D_{O\infty}$, to include the effects of thermal expansion due to heat release and will show that reduced parametric dependence of the solution on the effective Lewis number, characteristic of the thermodiffusive approximation, no longer holds. It should be noted here that the effects of thermal expansion has been studied previously~\cite{pearce2014taylor,daou2018taylor} in the context of steadily propagating premixed flames in narrow pipes and channels. The asymptotic solution for premixed flames was derived in these references in the limit where the flame thickness becomes very large in comparison with the pipe radius in which case the diffusive transport within the flame is essentially controlled by Taylor's dispersion processes with the result that the flame propagation speed depending on the flow properties. The solution method employed in \textbf{I} and in here follows similar approach except now that the thick-flame limit\red{~\cite{daou2002thick}} for the premixed flame is replaced by the thick mixing layer limit for the diffusion flame. Since the diffusion flame separates the fuel from the oxidizer, the effective Taylor diffusion coefficients are different on both sides of the reaction sheet if the molecular diffusion coefficients of the fuel and oxidizer are different; this aspect of the diffusion flame calls for a consideration of a thin region encompassing the reaction sheet, as already identified in \II.

\section{Preliminary considerations}

Firstly, we shall establish that even in the presence of thermal expansion, the prevailing pressure gradient acting on the mixing layer for $t^*\gg a^2/D_{O\infty}$ is still the imposed pressure gradient corresponding to the Poiseuille flow. If $x^*$ is the axial coordinate in the laboratory frame, at time $t^*$, the axial extent of the mixing layer is given by $z^*=x^*-Ut^*\sim \sqrt{D_{O\infty}t^*}$. In the absence of thermal expansion, the imposed pressure gradient is 
\begin{equation}
    \left(\frac{dp^*}{dz^*}\right)_i = \frac{8\mu U}{a^2} \label{poi}
\end{equation}
where $\mu$ is the constant viscosity. The thermal expansion due to heat release will also generate a pressure gradient in the axial direction. To estimate this, consider the following argument. In the moving frame, a point in the mixing layer will experience a density variation $\Delta \rho^*$, where $\Delta \rho^*$ is the density difference between the burnt and unburnt gas mixture, as a function of time. Similarly, the instantaneous density variations across the moving mixing layer will also be of the order $\Delta \rho^*$. If $U_e$ is the characteristic axial velocity in the moving frame induced by the expansion of the gas, then from the continuity equation, we have
\begin{equation}
    \frac{\Delta \rho^*}{t^*}\sim \frac{\Delta \rho^* U_e}{z^*}
\end{equation}
which implies that $U_e\sim \sqrt{D_{O\infty}/t^*}$. The pressure gradient associated with this axial motion is 
\begin{equation}
    \left(\pfr{p^*}{z^*}\right)_e \sim \frac{\rho_\infty U_e^2}{z^*} \sim \frac{\rho_\infty \sqrt{D_{O\infty}}}{t^*\sqrt{t^*}}. \label{exp}
\end{equation}

At times $t^*\sim a^2/D_{O\infty}$, equations~\eqref{poi} and~\eqref{exp} lead to
\begin{equation}
    \left(\frac{dp^*}{dz^*}\right)_i \sim \Pec \left(\pfr{p^*}{z^*}\right)_e
\end{equation}
where we used the relation $\mu=\rho^*\nu^*\sim\rho_\infty \nu_\infty\sim \rho_\infty D_{O\infty}$. This indicates that the pressure gradient developed by thermal expansion is comparable to the imposed pressure gradient for order unity Peclet number. Note that at these times, the mixing layer lies near the entrance region, where the streamwise pressure gradient is not known apriori even \red{when there is no} thermal expansion (a typical feature of developing internal flows). Indeed, Taylor's analysis of dispersion is not valid at these times and we do not address this problem here. On the other hand, when $t^*\gg a^2/D_{O\infty}$, we have 
\begin{equation}
    \left(\frac{dp^*}{dz^*}\right)_i \gg \Pec \left(\pfr{p^*}{z^*}\right)_e.
\end{equation}
Thus, in the first approximation, the pressure gradient acting on the mixing layer is constant and equal to the imposed pressure gradient \red{along} the pipe. This condition will be used in developing the solution for $t^*\gg a^2/D_{O\infty}$.

\section{Formulation}
Consider a one-step irreversible Arrhenius reaction
\begin{equation}
    \ch{F + s O_2 -> P + q}
\end{equation}
where $s$ \red{refers} to the mass of oxygen required to burn unit mass of fuel and $q$ \red{denotes} the amount of heat released per unit mass of fuel consumed. Let the fuel mass fraction $Y_F^*$ in the fuel stream be $Y_{F,\infty}$ and the oxidizer mass fraction $Y_O^*$ in the oxidizer stream be $Y_{O,\infty}$. The temperature of both \ared{streams} are equal, set to $T_\infty$, with corresponding equal densities of $\rho_\infty$ \ared{as follows from the low Mach number equation of state upon neglecting variations in the molecular weight}. The number of moles of fuel burnt per unit volume per unit time \ared{takes the} form $\omega^* = k Y_F^*Y_O^* k^*(T^*)$ \ared{involving the reaction-rate constant $k$ whose dependence on temperature and
pressure need not be specified in our Burke-Schumann analysis.}

Let the coordinates in the frame moving with the mean velocity $U$ be $\vect x^*=(z^*,r^*,\theta^*)$ with associated velocity components $\vect v^*=(u^*,v^*,0)$; far away from the flame surface, the longitudinal velocity component in the moving frame is given by $u^*=U(1-2r^{*2}/a^2)$.  We set the thermal diffusivity $D_T^*=D_O^*$, equivalent to setting the oxidizer Lewis number to unity. We shall also assume that $\rho^*\nu^*=\rho_\infty\nu_\infty$, $\rho^*D_F^*=\rho_\infty D_{F\infty}$ and $\rho^*D_O^*=\rho_\infty D_{O\infty}$. Introducing then the following non-dimensional variables and parameters
\begin{align}
    & \vect x=\frac{\vect x^*}{a}, \quad t = \frac{t^*}{a^2 /D_{O\infty}}, \quad \vect v=\frac{\vect v^*}{U}, \quad  p=\frac{p^*}{\rho_\infty \nu_\infty U /a}, \quad T=\frac{T^*}{T_\infty},\\ &\rho = \frac{\rho^*}{\rho_\infty}, \quad Y_F = \frac{Y_F^*}{Y_{F,\infty}}, \quad Y_O=\frac{Y_O^*}{Y_{O,\infty}}, \quad   \omega= \frac{\omega^*}{a^2/(\rho_\infty D_{O\infty})},\\ &S= \frac{s Y_{F,\infty}}{Y_{O,\infty}}, \quad Q= \frac{qY_{F,\infty}}{c_p T_\infty},\quad  \Pec = \frac{aU}{D_{O\infty}}, \quad \Lew = \frac{D_O^*}{D_F^*}, \quad \Pra = \frac{\nu^*}{D_O^*},
\end{align}
the governing equations in the low Mach-number approximation become
\begin{align}
    &\frac{D\rho}{Dt} + \Pec \rho \nabla\cdot\vect v = 0,\quad 
    \frac{\rho}{\Pra}\frac{Du}{Dt} = - \nabla p + \nabla^2 \vect v + \frac{1}{3}\nabla (\nabla\cdot\vect v), \\
    &\rho \frac{DY_F}{Dt} = \frac{1}{\Lew}\nabla^2 Y_F - \omega,\quad
    \rho \frac{DY_O}{Dt} = \nabla^2 Y_O - S\omega,\\
    &\rho \frac{DT}{Dt} = \nabla^2 T - Q\omega, \quad \rho T=1
\end{align}
where $D/Dt=\pfi{}{t}+\Pec \vect v\cdot\nabla$ is the material derivative, $\Lew$ is the Lewis of the fuel, $\Pra$ is the Prandtl number and $c_p$ is the specific heat at constant pressure of the gas mixture. Here $\omega= k(T)Y_FY_O$ is the non-dimensional reaction rate in which $k(T)$ is the non-dimensional rate constant.

\ared{We} consider the following initial condition to be imposed  at $t=0$,
\begin{equation}
    u=1-2r^2, \quad v=0, \quad T=1, \quad Y_F = 1-Y_O= \begin{cases}
    1, \quad \text{for}\,\, z<0,\\
    0, \quad \text{for}\,\, z>0.  \label{init}
    \end{cases}
\end{equation}
For $t>0$, we have, as $z\rightarrow \infty$,
\begin{equation}
    u = 1-2r^2, \quad v=0, \quad Y_F=0, \quad Y_O=1, \quad T=1
\end{equation}
and as $z\rightarrow -\infty$,
\begin{equation}
   u = 1-2r^2, \quad v=0, \quad Y_F=1, \quad Y_O=0, \quad T=1.
\end{equation}
Similarly at $r=0$, we have
\begin{equation}
    \pfr{u}{r}=0, \quad v=0, \quad \pfr{Y_F}{r}=0, \quad \pfr{Y_O}{r}=0, \quad \pfr{T}{r}=0
\end{equation}
and at $r=1$, we impose
\begin{equation}
     u=-1, \quad v=0, \quad \pfr{Y_F}{r}=0, \quad \pfr{Y_O}{r}=0, \quad \pfr{T}{r}=0
\end{equation}
\ared{corresponding} to impermeable, adiabatic walls.

At times $t\sim 1/\ep$ where $\ep\rightarrow 0$ is a small parameter, mixing affects a region of axial extent $z\sim 1/\sqrt\ep$ and \red{radial extent} $r\sim 1$. \ared{As such, we introduce the order unity variables}
\begin{equation}
    \tau = \ep t, \quad \xi = \sqrt\ep z.
\end{equation}
In these rescaled coordinates, the required equations become
\begin{align}
    &\ep \pfr{\rho}{\tau} + \Pec\left[\sqrt\ep\pfr{}{\xi}(\rho u) + \frac{1}{r}\pfr{}{r}(\rho rv)\right] = 0, \label{conmain}\\
    &\ep\frac{\rho}{\Pra} \pfr{u}{\tau} + \rho \frac{\Pec}{\Pra}\left(\sqrt\ep u\pfr{u}{\xi} + v\pfr{u}{r}\right) =-\sqrt\ep\pfr{p}{\xi} +  \frac{1}{r}\pfr{}{r}\left(r\pfr{u}{r}\right) +  \frac{4\ep}{3} \pfr{^2u}{\xi^2}  + \frac{\sqrt\ep}{3r} \pfr{^2(rv)}{\xi\partial r},\\
    &\ep\frac{\rho}{\Pra} \pfr{v}{\tau} + \rho \frac{\Pec}{\Pra}\left(\sqrt\ep u\pfr{v}{\xi} + v\pfr{v}{r}\right) =-\pfr{p}{r} +   \frac{4}{3r}\pfr{}{r}\left(r\pfr{v}{r}\right)  - \frac{4v}{3r^2}+  \ep \pfr{^2v}{\xi^2}   +  \frac{\sqrt\ep}{3} \pfr{^2 u}{\xi\partial r},\\
    &\ep  \rho \pfr{Y_F}{\tau} + \rho \Pec\left(\sqrt\ep u\pfr{Y_F}{\xi} + v\pfr{Y_F}{r}\right) = \frac{1}{\Lew}\left\{\frac{1}{r}\pfr{}{r}\left(r\pfr{Y_F}{r}\right) + \ep \pfr{^2Y_F}{\xi^2} \right\} - \omega ,\\
      &\ep \rho \pfr{Y_O}{\tau} + \rho \Pec\left(\sqrt \ep u\pfr{Y_O}{\xi} + v\pfr{Y_O}{r}\right) = \frac{1}{r}\pfr{}{r}\left(r\pfr{Y_O}{r}\right) + \ep \pfr{^2Y_O}{\xi^2} - S\omega , \label{Omain}\\
     &\ep \rho \pfr{T}{\tau} + \rho \Pec\left(\sqrt \ep u\pfr{T}{\xi} + v\pfr{T}{r}\right) = \frac{1}{r}\pfr{}{r}\left(r\pfr{T}{r}\right) + \ep \pfr{^2T}{\xi^2}  + Q \omega \label{Tmain}
\end{align}

An asymptotic solution to the foregoing equations will be derived in the limit $\ep\rightarrow 0$. We shall assume that the Burke-Schumann limit $k\rightarrow \infty$ is valid, \ared{leading to the chemical-equilibrium condition $Y_F Y_O = 0$}. In this limit, the \ared{reaction terms become} Dirac delta functions at the flame location. \ared{As shown in \textbf{I}, \red{in this limit} a multi-layered structure emerges \red{for the mixing layer} as $\epsilon \to 0$}. \ared{In the first approximation the flame appears as a plane sheet located at $\xi=\xi_f(\tau)$, separating the fuel and oxidizer. These regions of fuel and oxidizer, characterized by distances from the reaction sheet $|\xi - \xi_f(\tau)| \sim 1$ and concentrations $Y_F\sim Y_O \sim 1$, are referred to as the outer regions which flank a thin flame region, of axial extent $|\xi - \xi_f(\tau)| \sim \sqrt{\epsilon}$, (an axial distance on the order of the pipe radius) where one finds small reactant concentrations on the order of $Y_F\sim Y_O \sim \sqrt{\epsilon}$ }. In \red{the thin inner} region, the flame surface is no longer a plane sheet, but is determined by a nonlinear boundary value problem which describes the radial variations of the flame shape. When the Lewis number of the fuel is unity, there is an explicit expression for the flame shape. 
\ared{In addition to describing the flame shape, the inner region provides a jump condition necessary for completing the solution in the outer regions. }

\section{The equations of the outer region}
In the range outside the thin region $\xi-\xi_f(\tau)\sim \sqrt\ep$, either the fuel or the oxidizer is identically zero, as mentioned above. The equations \ared{for} the fuel and the oxidizer \ared{regions} can thus be derived separately.

\subsection{Fuel side, $-\infty<\xi<\xi_f(\tau)$}

On the fuel side of the reaction sheet, $Y_O=0$ so that equations~\eqref{conmain}-\eqref{Tmain} reduce to \ared{the reaction-free system}
\begin{align}
    \ep \pfr{\rho}{\tau} + \Pec\left[\sqrt\ep\pfr{}{\xi}(\rho u) + \frac{1}{r}\pfr{}{r}(\rho rv)\right] &= 0, \label{confu}\\
   \rho \frac{\Pec}{\Pra}\left(\sqrt\ep u\pfr{u}{\xi} + v\pfr{u}{r}\right) &=-\sqrt\ep\pfr{p}{\xi} + \frac{1}{r}\pfr{}{r}\left(r\pfr{u}{r}\right) + O(\ep),\\
   0 &=-\pfr{p}{r} +  \frac{4}{3r}\pfr{}{r}\left(r\pfr{v}{r}\right)  - \frac{4v}{3r^2} + O(\sqrt\ep),\\
    \ep  \rho \pfr{Y_F}{\tau} + \rho \Pec\left(\sqrt\ep u\pfr{Y_F}{\xi} + v\pfr{Y_F}{r}\right) &= \frac{1}{\Lew}\left\{\frac{1}{r}\pfr{}{r}\left(r\pfr{Y_F}{r}\right) + \ep \pfr{^2Y_F}{\xi^2} \right\},\\
     \ep \rho \pfr{T}{\tau} + \rho \Pec\left(\sqrt \ep u\pfr{T}{\xi} + v\pfr{T}{r}\right) &= \frac{1}{r}\pfr{}{r}\left(r\pfr{T}{r}\right) + \ep \pfr{^2T}{\xi^2} \label{Tfu}
\end{align}

For the asymptotic description of the solution with $\tau$ and $\xi$ are of order unity and $\epsilon \to 0$ we shall introduce
\begin{align}
    u&=u_0 + \sqrt\ep u_1 + \ep u_2 + \cdots, \label{uexp}\\
    v&=\sqrt\ep v_0 + \ep v_1 + \ep\sqrt\ep v_2 + \cdots,\\
    p&=p_0/\sqrt\ep +  p_1 + \sqrt \ep p_2 + \cdots,\\
    Y_F&= F_0 + \sqrt \ep F_1 + \ep F_2 + \cdots,\\
    T&= T_0 + \sqrt \ep T_1 + \ep T_2 + \cdots. \label{Texp}
\end{align}
\ared{The density} is given by $\rho=\rho_0 + \sqrt \ep \rho_1 + \cdots = 1/T_0 - \sqrt\ep T_1/T_0^2+\cdots$ \ared{as follows from the equation of state}. The expansion for $v$ follows from \red {the continuity equation} using the fact that the geometry we examine is slender because $z\sim 1/\sqrt \ep \gg r\sim 1$ and $u\sim 1$. The expansion for the pressure \ared{ensures that axial pressure gradients balance radial diffusion at leading order}. 

At leading order \ared{we have}
\begin{align}
    \pfr{}{\xi}(\rho_0u_0) + \frac{1}{r}\pfr{}{r} (\rho_0 r v_0) &=0, \label{con1}\\
    - \pfr{p_0}{\xi} + \frac{1}{r} \pfr{}{r} \left(r\pfr{u_0}{r}\right) &=0,\\
     \pfr{p_0}{r} &=0,\\
    \frac{1}{\Lew r}\pfr{}{r}\left(r\pfr{F_0}{r}\right)&=0,\\
    \frac{1}{ r}\pfr{}{r}\left(r\pfr{T_0}{r}\right) &=0.
\end{align}
Integrating these equations with appropriate boundary conditions including the condition that the pressure gradient is constant and equal to the imposed pressure gradient \ared{(see Section 2) gives}
\begin{align}
    u_0 = (1-2r^2),\quad
    v_0 = - \frac{1}{2\rho_0}\pfr{\rho_0}{\xi} (r-r^3),\quad
    \frac{dp_0}{d\xi}=-8, \label{v0} \\
     F_0=F_0(\xi,\tau), \quad T_0=T_0(\xi,\tau), \quad \rho_0=\rho_0(\xi,\tau).
\end{align}
\ared{The e}quations satisfied by the unknown functions $F_0$, $T_0$ and $\rho_0$ will be derived later.

At the next order, \ared{one obtains}
\begin{align}
    \pfr{\rho_0}{\tau} + \Pec \left[\pfr{}{\xi} (\rho_0 u_1 + \rho_1 u_0) + \frac{1}{r} \pfr{}{r} (\rho_0 rv_1 + \rho_1 r v_0)\right] =0, \label{con2} \\
    \frac{\Pec}{ \Pra}\rho_0 \left( u_0 \pfr{u_0}{\xi} + v_0 \pfr{u_0}{r}\right) = -\pfr{p_1}{\xi} + \frac{1}{r} \pfr{}{r} \left(r\pfr{u_1}{r}\right),\\
    0= - \pfr{p_1}{r},\\
    \Pec\rho_0  u_0 \pfr{F_0}{\xi}  = \frac{1}{\Lew r} \pfr{}{r}\left(r\pfr{F_1}{r}\right),\\
   \Pec\rho_0  u_0 \pfr{T_0}{\xi}  = \frac{1}{r} \pfr{}{r}\left(r\pfr{T_1}{r}\right).
\end{align}
These equations also can be integrated in a straightforward manner to obtain
\begin{align}
    p_1 &= p_1(\xi,\tau), \label{new1}\\
    u_1 &= \frac{1}{4}\pfr{p_1}{\xi}(r^2-1) + \frac{\Pec}{2\Pra} \pfr{\rho_0}{\xi}(r^4/4-r^6/9-5/36), \label{u1} \\
    F_1 &= F_{1a}(\xi,\tau) + \frac{\Pec\Lew}{4}\rho_0  \pfr{F_0}{\xi} (r^2-r^4/2) ,\\
    T_1 &= T_{1a}(\xi,\tau) + \frac{\Pec}{4}\rho_0  \pfr{T_0}{\xi} (r^2-r^4/2),\\
    \rho_1 &= - \rho^2 T_{1a} + \frac{\Pec}{4} \rho_0 \pfr{\rho_0}{\xi}(r^2-r^4/2) \label{new2}
\end{align}
\red{The unknown function $p_1(\xi , \tau)$ is obtained as follows. Multiplying \eqref{con2} by $2r dr$ and integrating with respect to $r$ from $0$ to $1$, we obtain the solvability condition}
\begin{equation}
    \pfr{\rho_0}{\tau}- \frac{\Pec}{8}\pfr{}{\xi}\left(\frac{\rho_0}{8}\pfr{p_1}{\xi}\right) -  \frac{\Pec^2}{48 } \frac{(2+\Pra)}{\Pra}\pfr{}{\xi}\left( \rho_0 \pfr{\rho_0}{\xi}\right) =0 \label{conf}
\end{equation}
\red{after some simplification using the solutions obtained so far.}
Further integration of this equation with respect to $\xi$ \red{and using the condition that the gradients vanish far away from the flame}, we find
\begin{equation}
    \frac{\Pec}{8}\pfr{p_1}{\xi} + \frac{\Pec^2}{48} \frac{(2+\Pra)}{\Pra}\pfr{\rho_0}{\xi} = \frac{1}{\rho_0}\int_{-\infty}^\xi\pfr{\rho_0}{\tau}d\xi. \label{dpdx}
\end{equation}
The pressure gradient $\pfi{p_1}{\xi}$ associated with the thermal expansion is thus solved once $\rho_0(\xi,\tau)$ is determined; $p_1$ is solved by integrating the above equation. 

At the next order, the equation for $Y_F$ and $T$ provides us
\begin{align}
   &\rho_0\pfr{F_0}{\tau} + \Pec \left[(\rho_1u_0 + \rho_0 u_1 )\pfr{F_0}{\xi} + \rho_0 u_0 \pfr{F_1}{\xi} + \rho_0 v_0 \pfr{F_1}{r}\right] = \frac{1}{\Lew}\left\{\frac{1}{r}\pfr{}{r}\left(r\pfr{F_2}{r}\right) + \pfr{^2F_0}{\xi^2}\right\},\\
    &\rho_0\pfr{T_0}{\tau} + \Pec \left[(\rho_1u_0 + \rho_0 u_1 )\pfr{T_0}{\xi} + \rho_0 u_0 \pfr{T_1}{\xi} + \rho_0 v_0 \pfr{T_1}{r}\right] = \frac{1}{r}\pfr{}{r}\left(r\pfr{T_2}{r}\right) + \pfr{^2T_0}{\xi^2}.
\end{align}
\ared{Averaging over $r$, and using \eqref{new1}--\eqref{new2} to simplify, one obtains a form similar to that of a one-dimensional unsteady planar mixing layer~\cite{lin1976asymptotic,peters2001turbulent}}
%Integrating the above two equations using the solvability condition, we reduce the system of equations, after some simplification and including the continuity equation~\eqref{conf}, to a form similar to that of the planar, unsteady mixing problem (cf.~\cite{lin1976asymptotic},~\cite{peters2001turbulent}),
\begin{align}
    \pfr{\rho_0}{\tau} + \pfr{}{\xi}(\rho_0u_e) & =0, \label{conff}\\
    \rho_0\pfr{T_0}{\tau}+ \rho_0 u_e \pfr{T_0}{\xi} & = \pfr{}{\xi}\left(\rho_0\mathcal{D}\pfr{T_0}{\xi}\right), \label{Tff}\\
    \rho_0\pfr{F_0}{\tau}+ \rho_0 u_e \pfr{F_0}{\xi} & = \frac{1}{\Lew}\pfr{}{\xi}\left(\rho_0\mathcal{D}_{F}\pfr{F_0}{\xi}\right) \label{Fff}
\end{align}
where we have introduced
\begin{equation}
    \rho_0 \mathcal{D} = 1+\frac{\Pec^2}{48}\rho_0^2\qquad \rho_0 \mathcal{D}_F = 1+ \frac{\Pec^2}{48}\Lew^2 \rho_0^2  \label{Taylor}
\end{equation}
and \ared{an ``effective" axial velocity induced by thermal} expansion\footnote[1]{Combining~\eqref{conff} and~\eqref{Tff} and using the relation $\rho_0T_0=1$ and boundary conditions, we can show that $u_e = \rho_0 \mathcal{D}\pfi{T_0}{\xi}$.}
\begin{equation}
    u_e = - \frac{\Pec}{8}\pfr{p_1}{\xi} -  \frac{\Pec^2}{48}\frac{(2+\Pra)}{\Pra}\pfr{\rho_0}{\xi}.
\end{equation}
\ared{The first equation follows directly from continuity \eqref{conf} written in terms of the effective velocity $u_e$}. Here $\mathcal{D}$ and $\mathcal{D}_{F}/\Lew$ \ared{may} \red{respectively} be considered as the Taylor diffusivities for \ared{the} oxidizer (or, temperature) and fuel, \red{measured in units of $D_{O\infty}$.} The \ared{diffusivities have been previously identified} for premixed flames~\cite{pearce2014taylor,daou2018taylor} where $\rho_0\mathcal{D}$ and $\rho_0\mathcal{D}_F/\Lew$ (i.e. including $\rho_0$) were termed as the Taylor diffusion coefficients. %, except that in their analysis the functions $\rho_0\mathcal{D}$ and $\rho_0\mathcal{D}_F/\Lew$ were termed as the Taylor diffusion coefficients. 
When $\Pec = 0$, the definitions \eqref{Taylor} imply that $\rho_0 \mathcal{D} = \rho_0 \mathcal{D}_F = 1$, consistent with the approximations introduced earlier in Section 3.

The continuity equation~\eqref{conff} can be eliminated by replacing $\xi$ with a Lagrangian coordinate
\begin{equation}
    \psi = \int_{0}^\xi \rho_0 \, d \xi.
\end{equation}
The energy equation~\eqref{Tff} and fuel mass-fraction equation~\eqref{Fff} become
\begin{align}
    \pfr{T_0}{\tau} & = \pfr{}{\psi}\left(\rho_0^2\mathcal{D}\pfr{T_0}{\psi}\right),\\
    \pfr{F_0}{\tau} & = \frac{1}{\Lew}\pfr{}{\psi}\left(\rho_0^2\mathcal{D}_F\pfr{F_0}{\psi}\right). \label{Foeq}
\end{align}
These are the required equations for $F_0(\psi,\tau)$ and $T_0(\psi,\tau)$. Since \red{the flame location} $\psi_f(\tau)$ (or, $\xi_f(\tau)$) has to be determined as a part of solution, we can introduce the following expansion 
\begin{equation}
    \psi_f(\tau) = \psi_{0,f}(\tau) + \sqrt \ep \psi_{1,f}+\cdots
\end{equation}
so that boundary conditions for \ared{the} aforementioned problem can be applied at $\psi_{0,f}$ in the first approximation. The boundary conditions for $\tau>0$ are
\begin{equation}
    F_0=T_0=1 \quad \text{as} \quad \psi\rightarrow -\infty \quad \text{and} \quad F_0= T_0-T_{0,f}=0 \quad \text{as} \quad \psi\rightarrow \psi_{0,f}^-
\end{equation}
where $T_{0,f}(\tau)$ is the leading-order flame temperature, \ared{to be determined as part of the solution as will be seen later}. The initial conditions are the same as in~\eqref{init}. Once $F_0(\psi,\tau)$ and $T_0(\psi,\tau)$ are solved, the velocity and pressure-gradient corrections $(\sqrt\ep u_1, \sqrt\ep v_0, \pfi{p_1}{\xi})$ to the Poiseuille flow can be calculated using~\eqref{v0},~\eqref{u1} and~\eqref{dpdx}.

\subsection{Oxidizer side, $\xi_f(\tau)<\xi<\infty$}
\ared{The solution for the oxidizer side proceeds in an analogous manner to that of the fuel side just described. Here the fuel is identically zero $Y_F = 0$ and the oxidizer is expanded as} 
\begin{equation}
    Y_O = O_0 + \sqrt\ep O_1 + \ep O_2 + \cdots.
\end{equation}
The development leads to 
\begin{align}
    \pfr{T_0}{\tau} & = \pfr{}{\psi}\left(\rho_0^2\mathcal{D}\pfr{T_0}{\psi}\right),\\
    \pfr{O_0}{\tau} & = \pfr{}{\psi}\left(\rho_0^2\mathcal{D}\pfr{O_0}{\psi}\right), \label{Ooeq}
\end{align}
with boundary conditions
\begin{equation}
    O_0=T_0=1 \quad \text{as} \quad \psi\rightarrow \infty \quad \text{and} \quad O_0= T_0-T_{0,f}=0 \quad \text{as} \quad \psi\rightarrow \psi_{0,f}^+
\end{equation}
and initial conditions given by~\eqref{init}.

\subsection{Enthalpy variable}

Since the governing equations for $Y_O$~\eqref{Omain} and $T$~\eqref{Tmain} are similar, we may introduce a conserved scalar
\begin{equation}
    H = \frac{1-T}{Q/S} + 1-Y_O \label{H}
\end{equation}
which satisfies a reaction-free transport equation similar to ~\eqref{Omain}. This enthalpy variable \cite{linan2017large} is defined such that $H(\psi\rightarrow-\infty,r,\tau) = 1$, $H(\psi\rightarrow\infty,r,\tau)= 0$ and $H(\psi,r,\tau\rightarrow 0)\rightarrow$ step function. \ared{Expanding as $H=H_0 + \sqrt \ep H_1 + \ep H_2 + \cdots$, it is straightforward to obtain at  leading order}
\begin{equation} \label{HOeq}
    \pfr{H_0}{\tau}  = \pfr{}{\psi}\left(\rho_0^2\mathcal{D}\pfr{H_0}{\psi}\right).
\end{equation}
\ared{The reaction free nature of $H$ ensures that this} equation is valid everywhere, i.e, for  $-\infty<\psi<\infty$. In place of the variable $T_0(\xi,\tau)$ whose derivative is discontinuous at $\psi_{0,f}$, we may solve for $H_0$ which has continuous derivative at the flame location.

\section{The equations of the inner region}

The inner region of thickness $O(\sqrt\ep)$ encompassing the flame surface is best described by the stretched coordinate $\eta=[\psi-\psi_f(\tau)]/\sqrt\ep$. In the inner region, the \red{departures of the independent variables from their values on the flame surface} are small, i.e. $Y_F\sim Y_O\sim T_f(\tau)-T\sim \rho-\rho_f(\tau)\sim \sqrt\ep$. \red{To facilitate the development,} we shall employ the Shvab-Zeldovich-Li{\~n}{\'a}n formulation~\cite{Linan1991structure} by introducing $X=(SY_F-Y_O)/\sqrt\ep$ and $\tilde X=(SY_F/\Lew - Y_O)/\sqrt\ep$ to describe the inner region, in which $\tilde X$ is a conserved scalar. Substituting the variables defined here into equations~\eqref{conmain}-\eqref{Omain} and collecting the leading-order terms result in
\begin{align}
    \rho_f\pfr{u}{\eta} + \frac{1}{r}\pfr{}{r}(rv) &= 0,\\
    \rho_f \frac{\Pec}{\Pra}\left( \rho_fu\pfr{u}{\eta} + v\pfr{u}{r}\right) &=-\rho_f\pfr{p}{\eta} + \frac{1}{r}\pfr{}{r}\left(r\pfr{u}{r}\right) +  \rho_f^2\pfr{^2u}{\eta^2} ,\\
     \rho_f\frac{\Pec}{\Pra}\left(\rho_f u\pfr{v}{\eta} + v\pfr{v}{r}\right) &=-\pfr{p}{r} + \frac{1}{r}\pfr{}{r}\left(r\pfr{v}{r}\right)  - \frac{v}{r^2}+  \rho_f^2\pfr{^2v}{\eta^2} ,\\
    \rho_f \frac{\Pec}{\Pra}\left(\rho_f u\pfr{X}{\eta} + v\pfr{X}{r}\right) &= \frac{1}{r}\pfr{}{r}\left(r\pfr{\Xt}{r}\right) +  \rho_f^2\pfr{^2\Xt}{\eta^2}
\end{align}
The equation for $H$ need not be written here as we can easily show that $H$ is conserved in the inner region. The quasi-steady incompressible Navier-Stokes equations must satisfy 
\begin{equation}
    u=1-2r^2, \quad v=0 \quad \text{at} \quad \eta=\pm\infty,
\end{equation}
consistent with the velocity field found outside the thin region at this order. These boundary conditions are also the solution so that
\begin{equation}
    u=1-2r^2, \quad v=0
\end{equation}
is valid everywhere in the inner region. We therefore have
\begin{equation}
    \rho_f^2\Pec(1-2r^2) \pfr{X}{\eta} = \frac{1}{r}\pfr{}{r}\left(r\pfr{\Xt}{r}\right) +  \rho_f^2\pfr{^2\Xt}{\eta^2} \label{Xeq}
\end{equation}
in which the function $X=X(\tilde X)$ can be determined from the equilibrium condition $Y_FY_O=0$ that prevails on the flame surface. \red{Averaging over $r$, we get the solvability condition}
\begin{align}
    \pfr{G}{\eta}=0 \quad \text{where} \quad G(\tau) = \int_0^1 2r\left[\Pec (1-2r^2)X - \pfr{\Xt}{\eta}\right] dr.
\end{align}
The flux $G$ thus remains constant across the inner region. If the flux is evaluated both as $\eta\rightarrow-\infty$ and as $\eta\rightarrow \infty$ using the two-term expansion derived in the outer region, we will find 
\begin{align}
    G(\tau) &= - \frac{S}{\Lew}\left(1+ \frac{\Pec^2}{48}\Lew^2\rho_{0,f}^2\right) \pfr{F_0}{\psi}\biggr\rvert_{\psi_{0,f}^-} = \left(1+ \frac{\Pec^2}{48}\rho_{0,f}^2\right) \pfr{O_0}{\psi}\biggr\rvert_{\psi_{0,f}^+}. \label{outerbc}
\end{align}
This relation provides the additional boundary condition required for the outer problem to determine $\psi_{0,f}(\tau)$. %Thus, we do not integrate this equation here since computational results are already presented in I for selected cases.

%\section{Solutions for the outer problem}
\section{The outer problem}
The basic equations and all the boundary conditions required to describe the outer structure are now available. The form of these equations and the initial and boundary conditions permit us to look for solutions that are functions only of $\chi=\psi/\sqrt{4\tau}$. In this self-similar coordinate, the flame location $\chi_{0,f}$ is a constant. 

The required equation and boundary conditions for $H_0$ are given by
\begin{equation}
    \frac{d}{d\chi}\left(\rho_0^2 \mathcal{D} \frac{dH_0}{d\chi}\right) + 2\chi \frac{dH_0}{d\chi}=0, \qquad H_0(-\infty)-1=H_0(\infty)=0 \label{H0}
\end{equation}
\ared{as follows from \eqref{HOeq}.}
On the fuel side $-\infty <\chi <\chi_{0,f}$, the problem of solving $F_0$ reduces to 
\begin{equation}
     \frac{d}{d\chi}\left(\rho_0^2 \mathcal{D}_F \frac{dF_0}{d\chi}\right) + 2\Lew \chi \frac{dF_0}{d\chi}=0, \qquad F_0(-\infty)-1=F_0(\chi_{0,f})=0,
\end{equation}
whereas on the oxidizer side $\chi_{0,f}<\chi<\infty$, we have
\begin{equation}
     \frac{d}{d\chi}\left(\rho_0^2 \mathcal{D} \frac{dO_0}{d\chi}\right) + 2\chi \frac{dO_0}{d\chi}=0, \qquad O_0(\chi_{0,f})=O_0(-\infty)-1=0,
\end{equation}
obtained from \eqref{Foeq} and \eqref{Ooeq}, respectively. The unknown constant $\chi_{0,f}$ is \ared{obtained} by using the condition~\eqref{outerbc}, that is, at $\chi=\chi_{0,f}$
\begin{equation}
    - \frac{S}{\Lew}\mathcal{D}_{F} \frac{dF_0}{d\chi} = \mathcal{D} \frac{dO_0}{d\chi}. \label{jump1}
\end{equation}
In these equations, the function $\rho_0$ appears through $\rho_0^2\mathcal{D}$ and $\rho_0^2\mathcal{D}_F$ \ared{each expressible in terms of $H_0$ and $O_o$ according to
\begin{align}
\rho_0^2\mathcal{D} & = [1+(1-O_0-H_0)Q/S]^{-1} + \frac{\Pec^2}{48} [1+(1-O_0-H_0)Q/S]^{-3}\label{Taylor_1}\\
\rho_0^2\mathcal{D}_F & = [1+(1-O_0-H_0)Q/S]^{-1} + \frac{\Pec^2}{48}\Lew^2 [1+(1-O_0-H_0)Q/S]^{-3}\label{Taylor_2}
\end{align}}
\ared{following from \eqref{Taylor} and \eqref{H}. These expressions take even simpler forms on the fuel side where $O_0 = 0$.} According to \eqref{H0}, since the function $\rho_0^2\mathcal{D}$ has a cusp at $\chi_{0,f}$ (associated with the jump in the temperature-gradient values across the reaction layer), $d^2H_0/d\chi_0^2$ will have a finite jump at that location, although $H_0$ and $dH_0/d\chi_0$ \red{are} continuous everywhere.

\ared{To summarize, the integration proceeds by solving for $F_0$ and $H_0$ in a fuel domain $-\infty \le \chi \leq \chi_{0,f}$ with $F_0 = H_0 = 1$ at $\chi = -\infty$ and $F_0 = H_0 - H_{0,f} = 0$ at $\chi = \chi_{0,f}$, where $H_{0,f}$ is the flame value of the enthalpy function (see below). Similarly one solves for $O_0$ and $H_0$ in the oxidizer domain $\chi_{0,f} \leq \chi \le \infty$ with $O_0 = H_0 - H_{0,f} = 0$ at $\chi = \chi_{0,f}$ and $O_0 -1 = H_0 = 0$ at $\chi = \infty$. The two unknown constants $\chi_{0,f}$ and $H_{0,f}$ are obtained in an iterative manner (e.g using Newton Raphson) once conditions \eqref{jump1} and $({\rm d} H_0/{\rm d}\chi)^+_f -({\rm d} H_0/{\rm d}\chi)^-_f = 0$ (expressing the continuous gradient of $H_0$ at the flame $\chi = \chi_{0,f}$) are satisfied. It is clear from \eqref{H0}--\eqref{Taylor_2} that the solution of the outer problem depends on four parameters, namely $S$, $Q/S$, $\Pec$ and $\Lew$. Sample solutions are described in the following section. }

\begin{figure}
% figure 2
\centering
\includegraphics[width=1\textwidth]{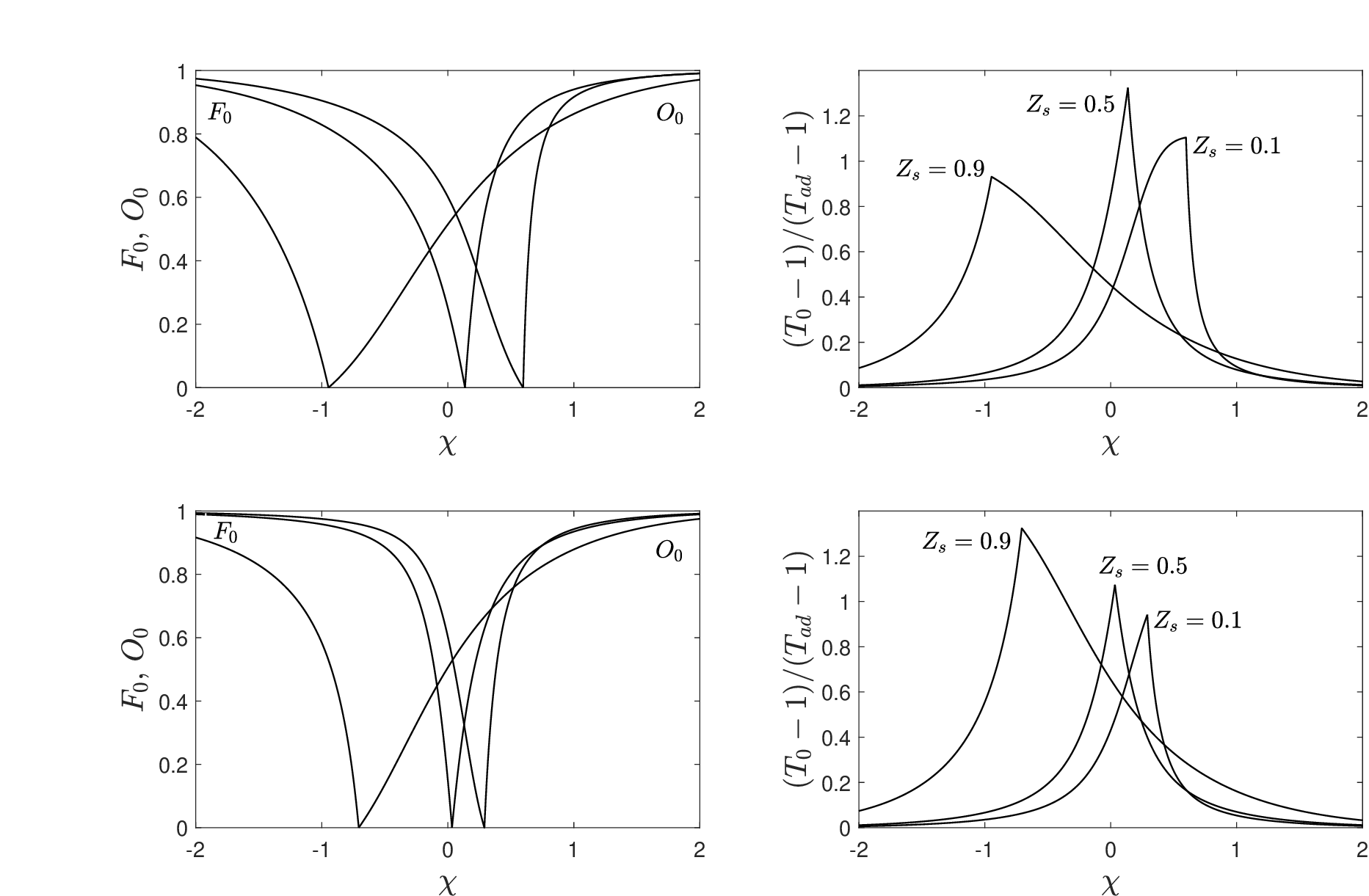}\\
\caption{The Burke-Schumann structure \ared{for $\Pec = Q/S = 10$ and $Z_s=(0.1,0.5,0.9)$. In the top two figures $\Lew = 0.3$ whereas in the bottom two $\Lew = 2.0$.} }
\label{fig:BS}
%\vspace{-0.10in}
\end{figure}

\begin{figure}[h!]
% figure 2
\centering
\includegraphics[width=1\textwidth]{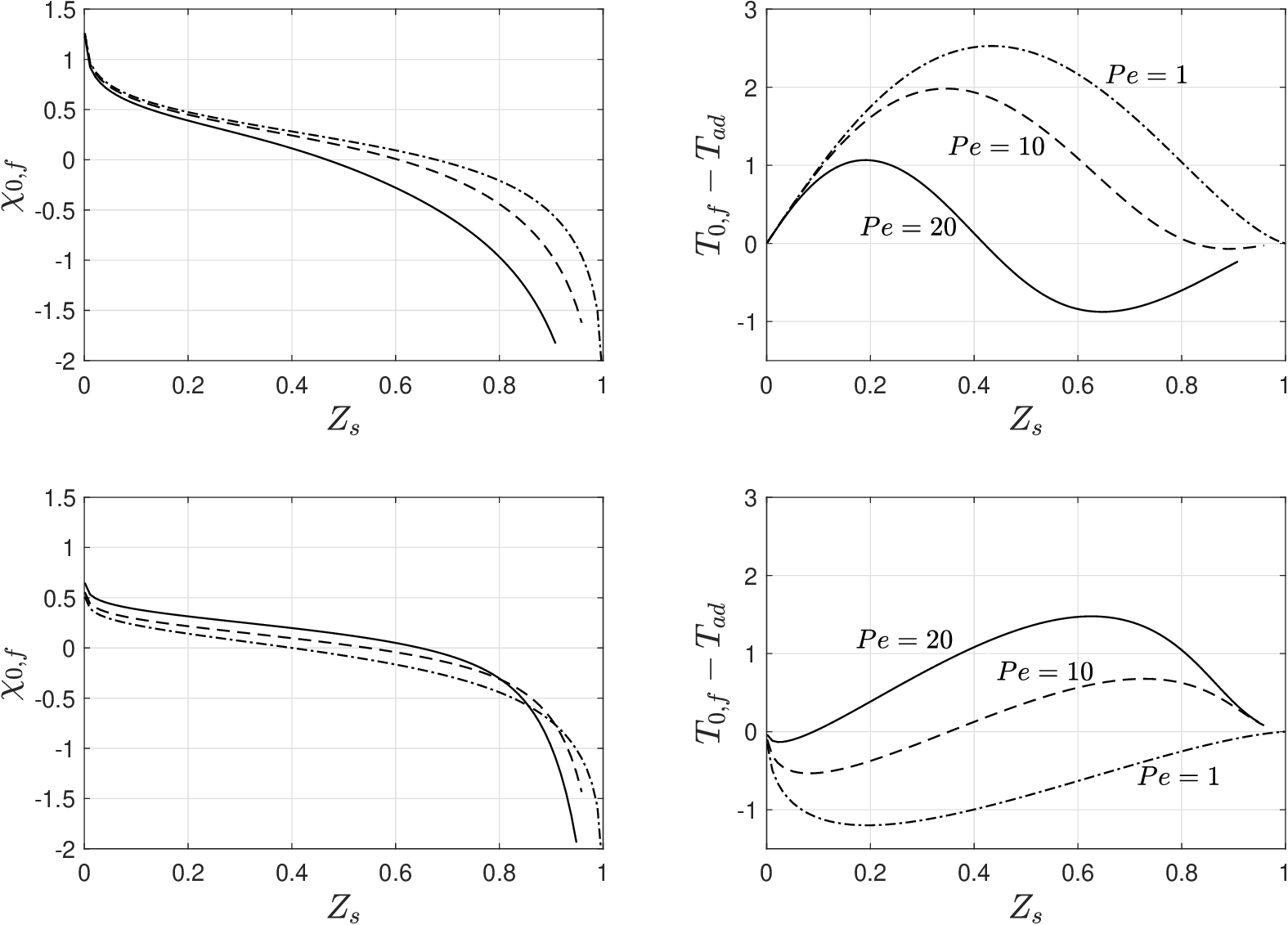}\\
\caption{The flame location and the flame temperature are described in the top two figures for $\Lew=0.3$ and the bottom ones for $\Lew=2.0$, both of which are calculated for $Q/S=10$ and for $\Pec=1$ (dot-dashed curves), $\Pec=10$ (dashed curves) and $\Pec=20$ (solid curves).}
\label{fig:fl}
%\vspace{-0.15in}
\end{figure}

\ared{
\section{Sample results and parametric dependence of outer problem}}

\ared{In presenting results, of the four controlling parameters} $S$, $Q/S$, $\Pec$ and $\Lew$, we shall fix $Q/S=10$ since this relation approximately holds for hydrocarbon-air and hydrogen-air combustion~\cite{linan2017large}. Values of \ared{the} stoichiometric mixture fraction $Z_s=1/(S+1)$ \ared{are} prescribed instead of $S$. Sample numerical results for three values of $Z_s=(0.1,0.5.0.9)$ and two different Lewis numbers $\Lew=0.3$ and $\Lew=2.0$, representing hydrogen and heavy hydrocarbon fuels, respectively, are shown in Fig.~\ref{fig:BS} \red{for $\Pec=10$}. The adiabatic flame temperature $T_{ad}=1+(1-Z_s)Q/S$ is used to \ared{scale} the temperature field while plotting the results. 

\red{When $\Pec=0$, the problem reduces to unsteady, planar diffusion flames which are mathematically identical to the steady, planar, counterflow diffusion flames. In such cases and with equal feed-stream temperatures~\cite{linan2017large}, fuels with $\Lew<1$ exhibit super-adiabatic flame temperatures for all values of $Z_s$ and vice versa for the case where $\Lew>1$. It is already shown in \textbf{I} that such observations are not to be expected in general, especially when the transport (Taylor) coefficients depend on the flow properties. In fact, the trend described above for $\Lew<1$ and $\Lew >1$ are shown to interchange (quite independently of $Z_s$) in the constant-density model once $\Pec$ becomes greater than $\sqrt{48/\Lew}$. Now observing the trend from the results shown in Fig.~\ref{fig:BS}, say for $\Lew=0.3$, we see that the flame temperature is super-adiabatic for $Z_s=0.1$ and $Z_s=0.5$ and sub-adiabatic for $Z_s=0.6$ when the Peclet number $\Pec=10$ still being less than $\sqrt{48/0.3}\approx 12.7$. Similarly, when $\Lew=2$, we see that the flame temperature is still sub-adibatic at $Z_s=0.1$ although the Peclet number $\Pec=10$ exceeded its transition point $\sqrt{48/4}\approx 4.9$. These results clearly indicate  1) that the transition between sub-adiabatic and super-adiabatic flame temperature does not occur at $\Pec=\sqrt{48/\Lew}$ and 2) that this transition does not occur for all values of $Z_s$ at the same time. To better understand these transitions, we plot the flame temperatures as functions of $Z_s$ for different values of $\Pec$.}

The flame location $\chi_{0,f}$ and the departure of the flame temperature 
\begin{equation}
    T_{0,f} = 1+\frac{Q}{S}(1-H_{0,f})
\end{equation}
from the adiabatic value are shown in Fig.~\ref{fig:fl}. \red{For the two values of $\Lew$ for which results are shown, it is clear from the figure that the Peclet number that causes the transition between super-adiabatic and sub-adiabatic not only depends on $\Lew$, but also on $Z_s$. The non-existence of a simple transition Peclet number (similar to that identified in \II) is due to} the fact that the Taylor diffusion coefficients~\eqref{Taylor} depend on the whole functional form of $\rho_0$ and not just on any particular value of it (say, the flame density).

%In the thermo-diffusive approximation, as shown in \II, the flame temperature transitions from sub-adiabatic to super-adiabatic for fuels with Lewis numbers greater than unity and the transition in the other way around for fuels with $\Lew <1$ for all values of $Z_s$ when $\Pec$ becomes greater than $\sqrt{48/\Lew}$. However, it is evident from the right-hand side plots of Figure.~\ref{fig:fl} that this kind of uniform transition does not exist when effects of thermal expansion is included. The fact that the Taylor diffusion coefficients~\eqref{Taylor} depend on the whole functional form of $\rho_0$ and not just on any particular value of it (say, the flame density) rules out the possibility of finding such simple transition points in variable-density models. 

 As $\Pec$ is increased, the flame temperatures of hydrogen-like fuels becomes sub-adiabatic initially for large values of $Z_s$ and becomes completely sub-adiabatic for all values of $Z_s$ when $\Pec$ is increased beyond a certain critical value. Similarly, the opposite phenomenon happens for fuels having $\Lew >1$. Both of these transitions as $\Pec$ is increased always initiate near the $Z_s=1$ boundary. \red{This is so because since at higher values of $Z_s$, the deficient reactant is the fuel where the coefficient $\mathcal{D}_F/\Lew$ being the controlling transport coefficient brings about the sub-adiabatic and super-adiabatic transition. On the contrary, these transitions, as $Z_s\rightarrow 0$, occur only at relatively larger values of $\Pec$ since the deficient reactant being the oxygen, whose Lewis number is taken here to be unity, has an opposing effect to such transitions.} As expected, whenever $\Lew=1$, the flame temperature is equal to $T_{ad}$.

%The mass of fuel consumed per unit area of flame surface per unit time $\dot m_F^*=-\rho_{0,f}^* \mathcal{D}_{F,f}^* \pfi{Y_F^*}{z^*}/\Lew$ written in the non-dimensional form 
%\begin{equation}
    %\dot m_F = \frac{\dot m_F^*}{\rho_\infty D_{O\infty} Y_{F,-\infty}/a} = -\frac{\sqrt\ep\rho_{0,f}^2 \mathcal{D}_{F,f}}{\sqrt{4\tau}\Lew} \frac{dF_0}{d\chi}\biggr\lvert_{\chi=\chi_{0,f}}=- \frac{(1+\rho_{0,f}^2\Lew^2\Pec^2/48)}{\sqrt{4t}\Lew} \frac{dF_0}{d\chi}\biggr\lvert_{\chi=\chi_{0,f}}.
%\end{equation}

%The effective Lewis number can be defined as
%\begin{equation}
   % \Lew_{\rm{eff}}= \frac{\mathcal{D}_f}{\mathcal{D}_{F,f}}\Lew= \frac{1+\rho_{0,f}^2\Pec^2/48}{1+\rho_{0,f}^2\Lew^2\Pec^2/48}\Lew.
%\end{equation}

\red{\section{The flame shape}}
\red{
 Once the outer problem is solved, the nonlinear boundary value problem~\eqref{Xeq} for $\tilde X(\eta,r,\tau)$ can be solved with the boundary conditions
 \beq
\pfr{ \tilde X}{ \eta} \rightarrow \frac{S}{\Lew} \left.\pfr{F_0}{ \psi}\right|_{\psi=\psi_{0,f}^-} \; {\rm as} \; \eta \rightarrow - \infty \quad {\rm and} \quad
\pfr{ \tilde X}{\eta} \rightarrow -\left.\pfr{ O_0}{ \psi}\right|_{\psi=\psi_{0,f}^+} \; {\rm as} \; \eta \rightarrow + \infty 
\eeq
obtained from matching with the outer expansions and $\pfi{\tilde X}{r}=0 $ at $r=0,1$. The gradients mentioned above are related by the flux $G(\tau)$, defined in~\eqref{outerbc}, carrying the quasi-steady time dependence of $\tilde X$. A time independent problem can be obtained with introduction of the rescaled dependent variables $\tilde \varphi = \tilde X/(G/\mathcal{D}_{F,f})$ and $\varphi = X/(G/\mathcal{D}_{F,f})$, where $\mathcal{D}_{F,f}$ is $\mathcal{D}_F$ [see \eqref{Taylor}] evaluated at the flame surface, functions of $r$ and the rescaled axial coordinate $\zeta = \eta/\rho_{0,f} = x - x_{0,f} + O(\sqrt{\epsilon})$.  In terms of these variables the problem reduces to
%\begin{equation}
%    \zeta = \frac{\eta}{\rho_{0,f}}, \quad \tilde \varphi = \frac{\tilde X}{G/ \mathcal{D}_{F,f}}, \quad  \varphi = \frac{X}{G/ \mathcal{D}_{F,f}}, \quad \mathcal{P} = \rho_{0,f}\Pec,
%\end{equation}
%where $\mathcal{D}_{F,f}$ is $\mathcal{D}_F$ (see \eqref{Taylor}) evaluated at the flame surface and $\mathcal{P}$ is a rescaled Peclet number. The correspondence between $\zeta$ and $x$ is $\zeta = x-x_{0,f} + O(\sqrt\epsilon)$. In terms of these variables we have
\begin{equation}
    \mathcal{P}(1-2 r^2) \pfr{\varphi}{\zeta}= \frac{1}{r}\pfr{}{r}\left(r\pfr{\tilde\varphi}{r}\right) + \pfr{^2\tilde\varphi}{\zeta^2}, \qquad
    \left\{\begin{array}{lll} \varphi=\Lew \tilde{\varphi} & {\rm for} & \tilde{\varphi}>0 \\
\varphi=\tilde{\varphi} & {\rm for} & \tilde{\varphi}<0 \end{array} 
\right., \label{varphieq}
\end{equation}
where $\mathcal{P} = \rho_{0,f} Pe$ is a rescaled Peclet number, to be integrated with boundary conditions
 \beq
\tilde{\varphi}(0,0)=0, \quad \pfr{ \tilde{\varphi}}{ r}(\zeta,0)=\pfr{ \tilde{\varphi}}{r}(\zeta,1)=0 %\label{bc_XXt}
\eeq
and
\beq
\pfr{ \tilde{\varphi}}{ \zeta} \rightarrow -1 \; {\rm as} \; \zeta \rightarrow - \infty \quad {\rm and} \quad \pfr{\tilde{\varphi}}{\zeta} \rightarrow - \frac{1+\Lew^2 \mathcal{P}^2/48}{1+\mathcal{P}^2/48} \; {\rm as} \;  \zeta \rightarrow + \infty. \label{bcvpt}
\eeq
The relation $\varphi=\varphi(\tilde \varphi)$ is obtained in a straightforward manner from the definitions of $X$ and $\tilde X$ using the equilibrium condition $Y_FY_O=0$. Note that the condition at the origin $\tilde \varphi(0,0)=0$ is introduced to eliminate the translational invariance of the problem. Thus the flame shape $\zeta_f=\zeta_f(r)$, identified from the condition $\tilde \varphi(\zeta_f,r)=0$ where $Y_F=Y_O=0$, passes through the origin. The flame shape thus determined from the time-independent problem above depends only on $\Lew$ and $\mathcal{P}$; the dependence of flame shape on $\mathcal{P}$ when $\Lew=1$ has been discussed previously~\cite{pearce2014taylor} for premixed flames.}

The nonlinear boundary problem~\eqref{varphieq}-\eqref{bcvpt} arising here is identical to that found in \textbf{I} for the constant-density model \red{with the Peclet number $\Pec$ replaced by the scaled Peclet number $\mathcal{P}$.  It is not hard to show with the help of the condition $\rho^* D_{O}^*=\rho_\infty D_{O\infty}$ introduced in section 2 that  $\mathcal{P}$ is just a Peclet number, but one that is defined with respect to the value of oxidizer diffusivity evaluated at the flame temperature. This allows results from the constant-density integrations of \textbf{I} to be applied here in the case of variable density in the following sense. The flame shapes calculated using the constant-density model will yield correct results if the constant density and diffusion coefficients evaluated at the flame temperature are selected as the representative constant values. However, the caveat with this approach is that unless $\Lew=1$ the flame temperature and consequently $\mathcal{P}$ are not known apriori. Thus the outer problem described in section 6 must be solved first before evaluating the flame shape.}

\red{When $\Lew=1$, $\mathcal{P}/\Pec = \rho_{0,f} = 1/[1+(1-Z_s)Q/S]$ which follows directly from the definition of $T_{ad}$. In this case, equations~\eqref{varphieq}-\eqref{bcvpt} admit the solution
\begin{equation}
    \tilde \varphi  + \zeta =- \mathcal{P}\left(\frac{r^2}{4}-\frac{r^4}{8}\right),
\end{equation}
for which the flame shape takes the simple form
\begin{equation}
    \zeta_f(r)=-\mathcal{P}\left(\frac{r^2}{4}-\frac{r^4}{8}\right).
\end{equation}
When $\Pec=0$, the solution is $\tilde\varphi=-\zeta$ which represents planar flames for arbitrary values of $\Lew$. In general, the flame shape depends on $Z_s$, $\Pec$ and $\Lew$ since $\rho_{0,f}$ depends on them. Regardless, the flame shapes predicted by~\eqref{varphieq}-\eqref{bcvpt} for various parametric values are similar to those identified in \textbf{I} and~\cite{pearce2014taylor}, except now that the axial extent of the flame for various cases are different. Thus, a better quantity to study the dependence of the three parameters is the maximum excursion of the flame shape. Since the flame shape always passes through the origin, the maximum excursion can simply be defined as $\zeta_{f,m}=-\zeta_f(1)$. If $\Lew=1$, we have
 \begin{equation}
     \zeta_{f,m} = \frac{\mathcal{P}}{8} = \frac{\Pec}{8[1+(1-Z_s)Q/S]}. \label{maxexc}
 \end{equation}
 }
 
 \begin{figure}[h!]
% figure 2
\centering
\includegraphics[width=0.6\textwidth]{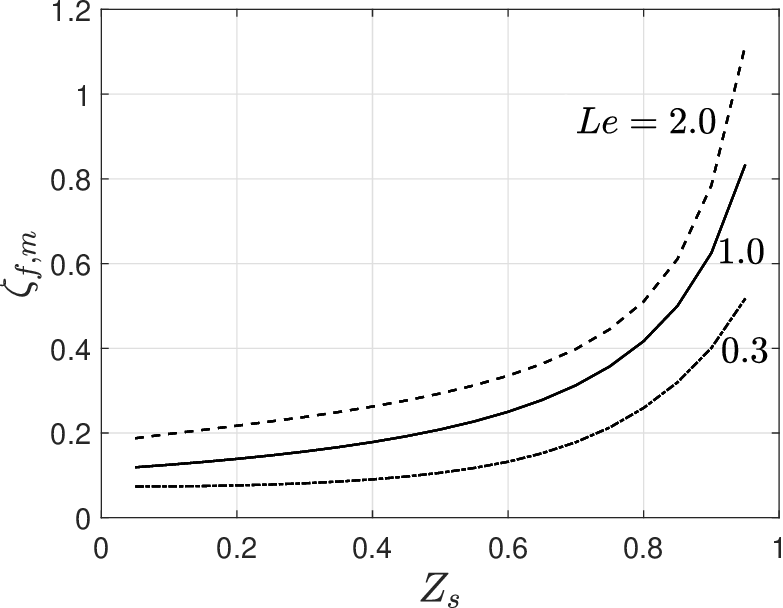}\\
\caption{Maximum excursions of the flame in the axial direction as functions of $Z_s$ for three different values of $\Lew=(0.3,1.0,2.0)$ and $\Pec=10$. }
\label{fig:max}
%\vspace{-0.15in}
\end{figure}

\red{We plot the maximum excursion in Fig.~\ref{fig:max} for three different values of $\Lew$ and $\Pec=10$ as a function of $Z_s$. Since the flame temperature $T_{0,f}$ is a decreasing function of $Z_s$, the flame density $\rho_{0,f}$ is an increasing function of $Z_s$. Thus for the case $\Lew=1$ we can see from~\eqref{maxexc} that the maximum excursion is an increasing function of $Z_s$. The same trend emerges for other values of $\Lew$, but not in the exact form given in~\eqref{maxexc}.}

%In an analogous fashion, for the constant density flames with $\Lew =1$ the maximum excursion occurs at $-\Pec/8$. As a result, since the flame density $\mathcal{P}/\Pec = \rho_{0,f} <1 $, the resulting flames in the variable density case are less``peaked" then their constant density versions. This result is expected to occur even when the Lewis number is non-unity.

\section{Concluding remarks}

The study initiated in \II, directed towards the understanding of diffusion flames in pipe flows with  applications in microcombustion devices, is extended here by including the effects of thermal expansion due to heat release. While the flow field induced by gas expansion is small, density variations associated with the expansion of the gas have significant influences on various quantities of interest such as the flame location and flame temperature of the diffusion flame.

The large-time asymptotic analysis showed that complicated partial differential equations corresponding to the variable-density model can be reduced to coupled nonlinear ordinary differential equations. These equations can be solved to describe the structure of the Burke-Schumann flame, thereby providing details about the flame location, flame temperature, fuel burning rate etc. However, the shape of the flame is still determined by the nonlinear partial differential equation~\eqref{Xeq} \ared{which was shown to reduce to a form equivalent to the constant density version by introduction of a rescaled Peclet number which accounts for the density variations at the flame}. It would be worthwhile in the future to investigate whether such methods can be extended to diffusion flames evolving in turbulent or pulsatile pipe flows \red{as well as under the influences of buoyancy forces. A study in the future concerning the stability of the Taylor-dispersion-controlled diffusion flames similar to that carried out for the premixed flames~\cite{daou2021effect}, especially in two-dimensional geometry, would also be beneficial.} Moreover, since aspects of heat transfer play a vital role in the design of microcombustion devices, it is of interest to see how the flame structure changes if the adiabatic walls are replaced by heat-conducting walls.

\section*{Acknowledgments}
The authors are grateful to Forman A. Williams and Antonio L. S{\'a}nchez for helpful discussion and suggestions. The authors also express thanks to Amable Li{\~n}{\'a}n for initiating this problem.

\bibliographystyle{tfq}
\bibliography{interacttfqsample}

\end{document}